\documentstyle[12pt,aaspp4,flushrt,epsf]{article}
\newcommand{\etal}{{\it et al.}}

\newcommand{\capj}{{\it Astrophys. J.}\ }
\newcommand{\mn}{{\it Mon. Not. R. Astron. Soc.}}
\newcommand{\gray}{$\gamma$-ray\ }
\newcommand{\grays}{$\gamma$-rays\ }

\begin{document}

\title{The Gamma-Ray Background from Blazars:  A New Look}

\author{F.W. Stecker\altaffilmark{1}}
\affil{Laboratory for High Energy Astrophysics\\
NASA Goddard Space Flight Center, Greenbelt, MD 20771}
\and
\author{M.H. Salamon}
\affil{High Energy Astrophysics Institute, Physics Department\\
University of Utah, Salt Lake City, UT 84112}
\altaffilmark{1}{also, High Energy Astrophysics Institute, Physics Department\\
University of Utah, Salt Lake City, UT 84112}

\begin{abstract}
We present the results of a new model calculation of
the \gray background produced by unresolved blazars, using the second
EGRET catalogue and taking account of flaring.  These results are compared
to the preliminary \gray background spectrum reported recently
by the EGRET team.  We find that blazars can account for the entire
extragalactic \gray background observed by EGRET.  In addition
the EGRET spectrum shows some 
evidence of a signature for the blazar
background, a concavity in the spectrum first pointed out in our earlier
paper.
\end{abstract}

\keywords{galaxies: active -- gamma-rays: theory -- quasars: general}

\section{Introduction}
\label{intro}

An isotropic, thus presumably extragalactic, component of the
cosmic \gray background above 30 MeV was first detected by the SAS-2 
satellite (\cite{fich78,thom82})
and has more recently been measured by the EGRET experiment on the Compton 
Gamma Ray Observatory (\cite{fich95}). The EGRET team has also
discovered that the only significant extragalactic point sources which 
produce \grays in this energy range are the catagory of objects called 
blazars (\cite{thom95,derm95a}).  These objects are active galactic 
nuclei whose jets are nearly pointing along our line of sight, resulting
in enhanced luminosity and variability within a narrow cone around the
jet axis, due to the highly relativistic motion of the source region
within the jet.
It can also be 
shown that physical mechanisms for producing diffuse radiation, such as the 
annihilation of antimatter or of supersymmetry dark matter particles, 
cannot produce an isotropic \gray flux above a few hundred MeV of
the magnitude observed (\cite{stec89}). Blazars would thus appear to be the 
only present candidates for producing the observed extragalactic
$\gamma$-radiation in this energy range.

With the new measurement of the extragalactic \gray radiation background
(EGRB) by EGRET (\cite{fich95}), and an enhanced blazar data set in the second EGRET
catalogue (\cite{thom95}), 
we have been led to reconsider an earlier model of \gray background
production by blazars (\cite{stec93,sala94}), which we have now expanded to 
include the effects of both flux {\it and} spectral variability of blazars
due to flaring.  A particularly important observation by EGRET is the apparent
hardening of spectra of blazars when they are in a flare state, which plays a
key role in our analysis.
Although this model is simplistic, it provides an excellent
fit to both the shape and amplitude of the EGRB, while simultaneously
accounting for both the number of blazars seen by EGRET, and their observed flux
distribution.

\section{The Basic Model}
\label{basic}

Our earlier model of \gray background production by blazars (\cite{stec93,sala94})
made the basic assumption that blazars seen in \grays above 100 MeV are
also seen in the radio (2.7 to 5.0 GHz) as flat spectrum radio quasars 
(FSRQs), with fewer blazars seen than FSRQs due to the existence of a narrower
beaming angle in \grays than in the radio (see also Dermer, 1995).  With the additional
assumption that the radio and \gray luminosity of these objects are linearly
related (see also Padovani, 1993),
one is able to use pre-existing radio luminosity functions (RLF) to
estimate the blazar \gray luminosity function (GLF), which can then be used to
calculate the contribution of unresolved blazars to the isotropic \gray
background.  A constraint on the model is that the number of {\it observed} blazars
in \grays estimated with this GLF correspond to the number actually observed
by EGRET, given their instrumental flux sensitivity.

We use here the same RLF as in our earlier work (\cite{dunl90}),
\begin{equation}
\rho_{r}(P_{r},z)=10^{-8.15}\left\{ \left[\frac{P_{r}}{P_{c}(z)}\right]^{0.83}
+\left[\frac{P_{r}}{P_{c}(z)}\right]^{1.96}\right\}^{-1},
\label{rlf.eq}
\end{equation}
where $\log_{10}P_{c}(z)=25.26+1.18z-0.28z^{2}$, and the units of source luminosity
$P_{r}$ and co-moving density $\rho_{r}$ are, respectively, 
W\,Hz$^{-1}$\,sr$^{-1}$ and Mpc$^{-3}\cdot$(unit interval of $\log_{10}P$)$^{-1}$.
$P_{r}$ is the differential radio luminosity (per steradian) measured at 2.7 GHz;
the RLF is constructed assuming pure luminosity evolution, with $\Omega_{0}=1.0$ and
$H_{0}=50$ km\,s$^{-1}$\,Mpc$^{-1}$.

The GLF is then given by 
\begin{equation}
\rho_{\gamma}(P_{\gamma f},z)=\eta\rho_{r}(P_{\gamma f}/\kappa,z),
\label{rhog.eq}
\end{equation}
where $\eta=(\theta_{\gamma}/\theta_{r})^{2}$ is the reduction of the number
of \gray blazars compared to FSRQs by virtue of a hypothesized smaller beaming angle
$\theta_{\gamma}$, and $\kappa$ is the proportionality constant
between the differential luminosities, $P_{\gamma f}=\kappa P_{r}$, which we
evaluate at the fiducial energies $E_{f}=100$ MeV and 
$\nu_{f}=2.7$ GHz.
The parameter $\kappa$ is estimated from the distribution of ratios of 
measured \gray and radio fluxes
of the EGRET blazars, where radio
fluxes at 2.7 GHz were obtained from \cite{wall85,kuhr81}, and \cite{ledd85}.

With this GLF, and appropriate values for $\eta$ and $\kappa$, an estimate
of the contribution to the EGRB from blazars was made (\cite{stec93,sala94})
and found to be a significant fraction of the measured EGRB.  In particular,
we noted that the spectrum of the EGRB due to blazars would be softer at
lower energies, and harder at higher energies,
a feature which appears to be
present in the most recent EGRB measurement (\cite{fich95}).
We also note that a recent
analysis using EGRET data to obtain an estimate of the 
GLF at the high end of the
blazar luminosity distribution is consistent with a luminosity and redshift
dependence of source distribution given by Eq.\ref{rlf.eq} (\cite{chia95}).

An obvious oversimplification of the earlier model was its lack of 
accounting for the well-known time variability of blazars, a
complex phenomenon which does not lead itself easily to simple
characterizations. It is known that
blazars can vary significantly over timescales of the order of a couple of
days or less (\cite{jang95,wagn95}).
Significant flaring in the GeV \gray\ range is perhaps best illustrated by
the intense flare which occurred during June 1991 in the source 3C279
(\cite{knif93}). Such flaring, in \grays as well as in other
regions of the electromagnetic spectrum, appears to be associated with
shocks propagating down relativistic jets and with the subsequent
acceleration of relativistic particles by these shocks (\cite{jang95,valt95}).

In this paper we incorporate this phenomenon in a simple
way, postulating the existence of two ``states'' of a blazar, a
``quiescent'' state and a ``flaring'' state, where the quiescent
blazar luminosity is amplified by a fixed factor $A$ during a flare. 
It is assumed that a blazar spends a fraction $\zeta$ of its time in a
flare state, where $\zeta \ll 1$.

\begin{figure}[htb]
\epsfxsize=8cm
\epsfbox{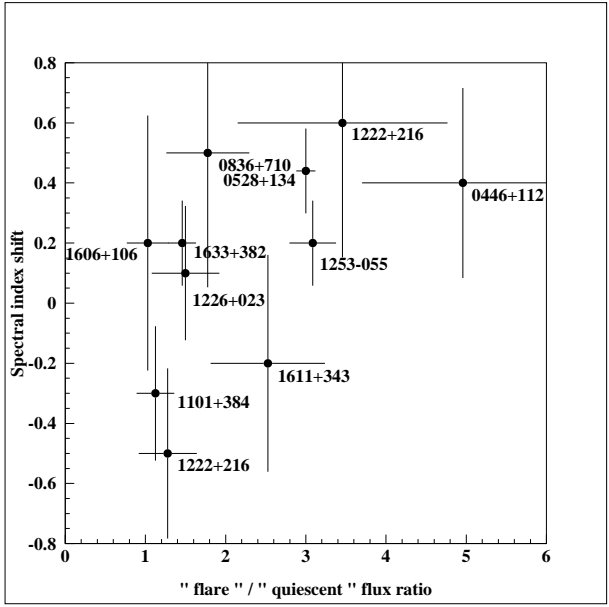}
\caption{A plot of the shift in measured blazar spectral index 
versus flux ratio for various blazars observed by ERGET during different 
viewing periods.  The flux values used in the ratios were those obtained for
the viewing periods in which spectral indices are determined.  The datum for
0528+134 is taken from Mukherjee \etal (1995).}
\label{fig1}
\end{figure}

In addition, the distribution of spectral indices of flare-state
blazars is taken to be harder than that of quiescent-state blazars (\cite{sree95}).
This is supported by EGRET's observations of flares in which the
hardening of the spectral index is statistically significant
{\it e.g.}, 0528+134 (\cite{mukh95}), and also as a statistical
trend as seen in Figure 1.  Here we have used the $2^{nd}$ EGRET catalogue
(\cite{thom95}) and considered all blazars for which more than one spectral
index is given (corresponding to different viewing periods).
For each object, the ratio of the integral \gray fluxes of the two
different periods is plotted against the spectral index shift, where
we construct the flux ratio so that it is always greater than unity.
A suggestive trend is
apparent, indicating a typical hardening of the spectral index by
$\sim 0.3$ during strong ($A>3$) flares.

The spectral index $\alpha$ is defined
by $E\frac{dF}{dE}(E)\propto E^{-\alpha}$, where $\frac{dF}{dE}(E)$ 
is the differential number flux
of \grays at energy $E$; equivalently, the \gray differential
luminosity at the source varies
with \gray energy as $P_{\gamma}(E)=P_{\gamma f}(E/E_{f})^{-\alpha}$,
consistent with all observed blazar \gray spectra.
The blazars' distribution of spectral 
indices $\alpha$ is determined from the $2^{nd}$ EGRET catalogue,
\begin{equation}
p(\alpha)=\frac{1}{N}\sum_{i=1}^{N}\frac{1}{\sigma_{i}\sqrt{2\pi}}
e^{-(\alpha-\alpha_{i})^{2}/2\sigma_{i}^{2}},
\label{pa.eq}
\end{equation}
where $\alpha_{i}$ and $\sigma_{i}$ are the catalogue's spectral index and
error for the $i^{\rm th}$ of N blazars.
We simply shift this distribution to obtain the separate distributions for the
flare and quiescent states: $p^{f}(\alpha)=p(\alpha-\Delta\alpha_{f})$, and
$p^{q}(\alpha)=p(\alpha-\Delta\alpha_{q})$, where we take
$\Delta\alpha_{q}=0.20$ and $\Delta\alpha_{f}=-0.05$, as discussed below.
As there is no evidence for a change in the blazar spectral index distribution
with redshift $z$ in the EGRET catalogue, we assume none.

Our GLFs for the two states of blazars, quiescent and flaring,
is a modification of Eq.\ref{rhog.eq},
\begin{equation}
\rho_{\gamma}(P_{\gamma f},z)=
(1-\zeta)\eta\rho_{r}(\frac{P_{\gamma f}}{\kappa},z) +
\zeta\eta\rho_{r}(\frac{P_{\gamma f}}{A\kappa},z),
\label{newrho.eq}
\end{equation}
where the quiescent and flare \gray luminosities, $P^{q}_{\gamma}$ 
and $P^{f}_{\gamma}$, are related to the radio luminosity as
$P^{q}_{\gamma}=\kappa P_{r}$ and $P^{f}_{\gamma}=A\kappa P_{r}$.
We note that 
although variability is seen in the radio at higher frequencies,
very little is observed at 2.7 GHz, the frequency at which our chosen
RLF is determined (\cite{reic93,valt95}).  Thus we need not be concerned with
the effects of variability in our reference RLF.

\section{The Observed Flux Distribution}
\label{flux}

For a blazar at redshift $z$ with luminosity $P_{\gamma f}$ (which is
the luminosity evaluated at the fiducial energy $E_{f}$ {\it at the source})
one can show that the integral flux $F(E_{f})$ {\it at Earth} is given by
\begin{equation}
F(E_{f})=P_{\gamma f}/\left[ \alpha(1+z)^{\alpha +1}r^{2}R_{0}^{2}\right],
\label{flux.eq}
\end{equation}
where $r$ is the co-moving coordinate of the blazar, and $R_{0}$ is the
present cosmological scale factor, 
$R_{0}r=(2c/H_{0})\left[1-(1+z)^{-1/2}\right]$
for $\Omega_{0}=1$ and $\Lambda=0$.  (The absence of $4\pi$ in the denominator
is because $P_{\gamma f}$ is a differential luminosity {\it per steradian}.)

The number of sources at redshift $z$ seen at Earth with an integral flux $F$
is given by
\begin{equation}
\frac{dN}{dF\,dz}\Delta z\,\Delta F =
4\pi R_{0}^{3}r^{2}\,\Delta r\,\rho_{\gamma}
(P_{\gamma f},z)\Delta(\log_{10}P_{\gamma f}),
\label{dndfdz.eq}
\end{equation}
where the relation between $F$ and $P_{\gamma f}$ is fixed by Eq.\ref{flux.eq}.
With this, the number of sources present per unit logarithm of flux $F$ is
found by integrating Eq.\ref{dndfdz.eq} over redshift,
\begin{equation}
\frac{dN}{d(\log_{10}F)}=\frac{16\pi c^{3}}{H_{0}^{3}}
\int_{0}^{z_{\rm max}}dz\,\frac{\left[(1+z)^{1/2}-1\right]^{2}}{(1+z)^{5/2}}
\left\{\eta(1-\zeta)\rho_{r}\left[\frac{P_{\gamma f}}{\kappa},z\right] +
\eta\zeta\rho_{r}\left[\frac{P_{\gamma f}}{A\kappa},z\right]\right\},
\label{dndf.eq}
\end{equation}
where we take $z_{\rm max}=5$.

For the choice of parameters discussed in Sec.\ref{egrb} below, 
the number of blazars in the sky as a function of integrated flux at Earth
is shown in Figure 2.  Here we show the number seen as 
quiescent-state blazars, and the number seen as flare-state blazars, and their
total.  Note that the two separate contributions represent the same underlying
(radio) distribution, shifted along the abscissa by the flare factor $A$
($A=5$), and along the ordinate by the flare duty cycle, $\zeta$
($\zeta=0.03$).  Also shown on the figure are the number of blazars seen
by EGRET as a function of flux $F(E_{f}=100$ MeV), binned logarithmically.
We take as our sample the 50 blazars identified in the 
2nd EGRET catalogue, along with 14 unidentified EGRET 
sources above 30 degrees galactic latitude which are most likely to be 
blazars (\cite{mukh94}).  The agreement is seen to be quite good, both in
the shape and amplitude of the experimental distribution.  The dropoff of data
below $F=2\times 10^{-7}$ cm$^{-2}$-s$^{-1}$ is due to the loss of 
blazar detection
efficiency near EGRET's minimum flux sensitivity of $\sim 1\times 10^{-7}$
cm$^{-2}$-s$^{-1}$.  In addition, there is incompleteness in the sky survey
at this sensitivity near the galactic plane, owing to the enhanced galactic
foreground emission.

\begin{figure}[htb]
\epsfxsize=8cm
\epsfbox{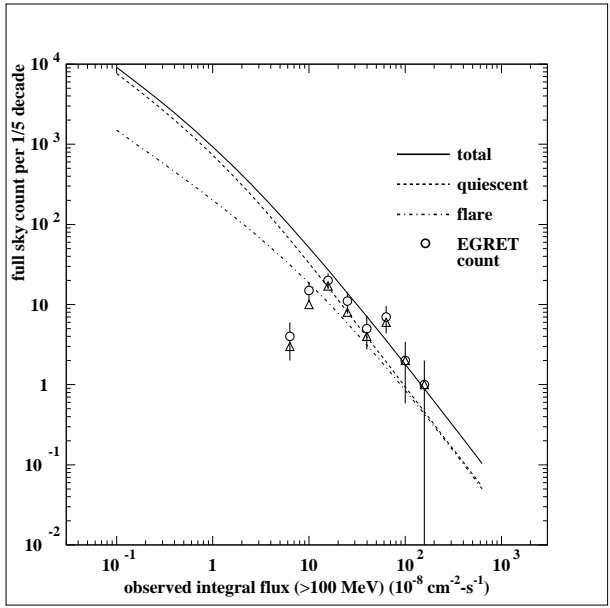}
\caption{The the predicted distibution of sources at various flux levels 
using the model parameters described in the text.
The open circles and their statistical error bars represent the 50 objects
identified as blazars in the second EGRET catalogue, plus the 14 unidentified,
high-latitude sources which are presumably blazars as well.  The open triangles
(no error bars) show only the 50 identified blazars' count per logarithmic bin.
Note that the ordinate is the
full sky count per {\it one-fifth} decade of flux.}
\label{fig2}
\end{figure}

The curves in Figure 2 give the number of sources {\it vs.} flux predicted
{\it at a single point in time}.  At any one point in time, a certain fraction
of the observed blazars will be in a flare state.  At another point in time,
however, a different set of blazars will be in a flare state.  Since EGRET
has viewed the entire sky multiple times (although not uniformly after its
inital full sky map), one expects that the number of flaring blazars seen by
EGRET would be greater than that indicated by the flare curve of Figure 2.
How much this should increase the contribution of
the flare component to the {\it total} EGRET blazar count is not clear, as many of
the flaring blazars are also detected in the quiescent state (\cite{mont95}).
Complicating this issue is the fact that many of the
multiple observations are made significantly off EGRET's vertical axis, 
increasing the minimum
flux sensitivity for detection.  The mean number of observations of an object
(with and without detections) is $\sim 7$, at various levels of sensitivity.
Were we to increase the flare contribution to the total EGRET count by a factor
of 3 to account for multiple viewings, we would still be reasonably consistent
with the data (allowing for off-axis inefficiencies at the lower flux levels).

An important point is that the distribution of spectral indices $\alpha$
measured by
EGRET contains those from both the quiescent and flare state populations, 
with the flare states contributing heavily to the EGRET distribution, as 
Figure 2 and the above discussion indicates.  Therefore the quiescent
state population, which we show below dominates the extragalactic
\gray background radiation, has a {\it softer} distribution of spectral indices
than that seen by EGRET.  

\section{The Extragalactic Gamma-Ray background}
\label{egrb}

Unresolved blazars, {\it i.e.}, those which are below detection threshold,
will contribute to a diffuse, isotropic background of \gray radiation.
We now consider whether unresolved blazars may be responsible for the
{\it entire} observed extragalactic \gray background.

The differential number flux $\frac{dF}{dE}^{(1)}$ of \grays 
received at Earth from a {\it single} source at redshift $z$ is given by
\cite{sala94}
\begin{equation}
\frac{dF}{dE}^{(1)}(E_{0},z)= 
\frac{H_{0}^{2}P_{\gamma f}(E_{0}/E_{f})^{-(\alpha +1)}}
{4c^{2}E_{f}(1+z)^{\alpha+1}\left[1-(1+z)^{-1/2}\right]^{2}},
\label{dfde1.eq}
\end{equation}
where $E_{0}$ is the \gray energy at Earth, $E_{f}$ is the fiducial
energy (100 MeV), and $\alpha$ is the spectral index of the single source.
We note that we neglect the minor effect of \gray extinction off the IR-to-UV diffuse
background (\cite{stec92}) for our energy region of interest, 0.1 to 100 GeV.

The total differential flux from unresolved blazars at a redshift $z$ is
found by integrating Eq.\ref{dfde1.eq} over the GLF $\rho_{\gamma}(P_{\gamma f},z)$:
\begin{equation}
\frac{dF}{dE}(E_{0})=4\pi R_{0}^{3}r^{2}\,dr\,\int d\alpha\,p(\alpha)
\int_{P_{\gamma f, {\rm min}}}^{P_{\gamma f, {\rm max}}}
\frac{dF}{dE}^{(1)}(E_{0},z)\rho_{\gamma}(P_{\gamma f},z)\,d(\log_{10}P_{\gamma f}),
\label{dfdez.eq}
\end{equation}
where $P_{\gamma f,{\rm max}}$ is the luminosity of a blazar at redshift $z$
which is just below the threshold for detection (found by subtituting EGRET's
integral flux sensitivity $F^{\rm EGRET}_{\rm min}(100$ MeV) 
into Eq.\ref{flux.eq}), and an integration
over the blazar spectral index distribution is made.  

We next integrate over redshift to obtain the total diffential flux at Earth 
at energy $E_{0}$.  We substitute Eq.\ref{newrho.eq} into Eq.\ref{dfdez.eq},
expressing the luminosity integral as one over the RLF (Eq.\ref{rlf.eq}),
obtaining
\begin{eqnarray}
 & \frac{dF}{dE}(E_{0})=\frac{c}{H_{0}E_{f}\,\ln10}\left\{
\int p^{q}(\alpha)\left(\frac{E_{0}}{E_{f}}\right)^{(\alpha +1)}
\int_{0}^{z_{\rm max}}\frac{dz}{(1+z)^{\alpha + 5/2}}
\int_{P_{r,{\rm min}}}^{P_{r,{\rm max}}^{q}(z)}
(1-\zeta)\eta\rho(P_{r},z)\,dP_{r} \right. & \nonumber \\
 & + \left. \int p^{f}(\alpha)\left(\frac{E_{0}}{E_{f}}\right)^{(\alpha +1)}
\int_{0}^{z_{\rm max}}\frac{dz}{(1+z)^{\alpha + 5/2}}
\int_{P_{r,{\rm min}}}^{P_{r,{\rm max}}^{f}(z)}\zeta\eta\rho(P_{r},z)\,dP_{r}
\right\}. & 
\label{totflux.eq}
\end{eqnarray}
The radio luminosity limits are given by 
$P_{r,{\rm max}}^{q}(z)=P_{\gamma f,{\rm max}}(z)/\kappa$, and
$P_{r,{\rm max}}^{f}(z)=P_{\gamma f,{\rm max}}(z)/A\kappa$.

This equation gives the results shown in Figure 3, plotted as $E^{2}\frac{dF}{dE}$
(the ``energy flux'') 
{\it vs.} $E$, for the parameter values $\kappa=4\times 10^{-11}$,
$\eta=1.0$, $\zeta=0.03$, $A=5$, $\Delta\alpha_{q}=0.20$, and
$\Delta\alpha_{f}=-0.05$.  Both the amplitude and shape of the calculated
energy flux
$E^{2}\frac{dF}{dE}$ matches that of the data, with the exception of the
two end data points, which have large systematic uncertainties (not shown)
(P. Sreekumar, personal communication, 1995).  
We note in particular the role of the spectral index shift
between the quiescent and flare state populations:  if Eq.\ref{totflux.eq}
were to be integrated over $\alpha$ using just EGRET's observed spectral
index distribution, the minimum in $E^{2}\frac{dF}{dE}$ would displaced to
a much lower energy, in contradiction to the data.

\begin{figure}[htb]
\epsfxsize=8cm
\epsfbox{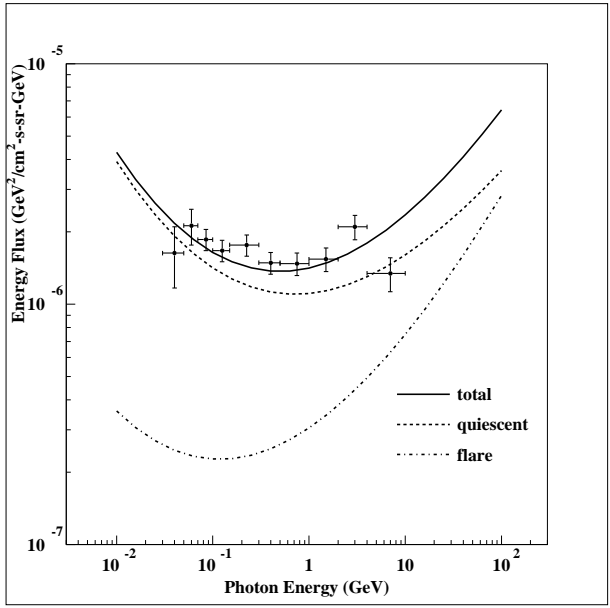}
\caption{The predicted background spectra from summing unresolved 
blazars compared with the preliminary EGRET data on the cosmic \gray\ 
background. The lower, middle, and upper curves show the 
contribution to the EGRB from flaring sources, 
from quiescent sources, and the their total.
The curves shown are for the parameters indicated in the text, which are 
the same as those used to generate Figure 2.}
\label{fig3}
\end{figure}

Thus we find that in order to match the observed spectral shape of the EGRB,
we {\it must} invoke a hardening of the flare-state spectra with respect to
the quiescent-state spectra, a feature which is already suggested by EGRET
observation (Figure 1).  In addition, we find that the {\it entire}
EGRB can be explained as being due to the \gray emissions of unresolved
blazars.

\section{Discussion}
\label{discuss}

We find that those unresolved blazars which make up the \gray background 
are primarily in the quiescent state, and therefore have a mixture of spectral indices 
shifted somewhat to the steep side (higher $\alpha$) than those
which make up the set of {\it observed} blazars. This is because the observed
blazars contain a higher mixture of flaring sources than those which make
up the background, owing to the high EGRET detector threshold and the 
shape of the high end of the GLF (Figure 2).  Thus the EGRET blazar
spectral index distribution will be intermediate between that of the
flare and quiescent populations, with the mean quiescent-state index being
$\Delta\alpha_{q}=0.20$ higher than the observed EGRET distribution, and
the mean flare-state index being $|\Delta\alpha_{f}|=0.05$ lower than
the EGRET distribution.  The difference in mean spectral index between the
flare and quiescent states, 0.25, is consistent with the trend indicated in
Figure 1.  In order to match the concavity of the observed EGRB spectrum,
we require a spectral index distribution that is softer than that seen by
EGRET, a condition that is met by this two-population model.
We note that a 
general feature of models which posit that the EGRB is due solely to blazars
is that the predicted EGRB spectra is {\it concave} and not a simple power law.
This is due to the EGRB being a superposition of blazar power-law spectra of
different spectral indices, where the softer
components dominate at lower energies, and harder components dominate at the
higher energies.  

The calculated EGRB spectrum in Figure 3 is shown down 
to an energy of 10 MeV. Of the few EGRET sources which have been detected 
by OSSE and COMPTEL below this energy, 7 by COMPTEL and 5 by OSSE 
(\cite{mcna95} and references therein), almost all show a 
flattening below $\sim 10$
MeV. There are not enough sources detected at the lower \gray energies to
calculate a background, although it is reasonable to assume that the
background from unresolved blazars should also flatten below 10 MeV. 
Note also that the calculated
energy flux of the EGRB increases
by more than a factor of 2 as one goes 
from 30 MeV 
down to 10 MeV energy. A preliminary analysis of COMPTEL data (\cite{kapp95})
gives a value 
for the GRB at 10 MeV which is more than a factor of 2 below our results. 
It is possible that the discrepancy here may be due to an
oversubtraction of detector backgrounds in obtaining the residual GRB from 
the COMPTEL data, this analysis also having given much lower values than 
those obtained by APOLLO 15 (\cite{trom77}) at lower energies.

Finally, we discuss the parameter values chosen to match the EGRET data.
A single flaring factor of $A=5$ is a gross simplification to what is
obviously a continuous distribution of flaring factors, and is perhaps as
large a value as one should take in this model, although a large $A$ helps
distinguish the separate roles of the two blazar states.  A surprising
result of this calculation is that we were driven to a large value of
$\eta$, which is the ratio of solid angles of the beamed \gray emission
to the beamed radio emission; in fact, our optimal value is unity.  This
contrasts with the results of our earlier work (\cite{sala94}) and the
model of Dermer (1995), where the emitted \grays are
emitted into a much smaller solid angle around the jet axis compared
to the beamed radio emission.  This smaller \gray beaming angle was invoked
in our earlier model
to explain geometrically why fewer blazars are seen 
in \grays than in the radio.  However, in the present 
model, the reason fewer \gray blazars
are seen is because the ratio $\kappa$ between the \gray and radio
luminosities is lower.

The relationship between the parameters $\eta$ and $\kappa$ can be seen
in Eq.\ref{dndf.eq}.  To match the observed number of sources per
logarithmic bin of flux, the quantity $\eta\kappa^{\Gamma}$ is fixed,
where $\Gamma$ is the local power law of the radio luminosity distribution,
$\rho_{r}(P_{r})\propto P_{r}^{-\Gamma}$.
When we include the effect of flare amplification, $\kappa$ is lowered
relative to our earlier model; thus $\eta$ becomes larger than in our
earlier model.
Owing to the
fact that we have introduced {\it two} states of blazar activity, we have
now have the freedom to raise $\eta$ so as to match the amplitude of the
EGRB, while simultaneously respecting the observed number of sources.
An indication that this added degree of freedom is not totally {\it ad hoc}
is seen by the fact that we {\it require} a difference in the mean spectral
indices of our two blazar populations which is consistent with that
suggested by EGRET observations.

Our conclusion then is that both the amplitude and shape of the extragalactic
\gray background spectrum as measured by EGRET
can be explained as being due solely to the
\gray emissions of unresolved blazars, if one assumes a simple relationship
between the \gray and radio luminosities of these objects.  A critical feature
of this simple model is that at a given time, a blazar is in one of two states: a
quiescent state or a flare state.

\acknowledgements
We thank P. Sreekumar for many helpful conversations and criticisms of
the manuscript.


\begin{thebibliography}{DUM}
\bibitem[Becker \etal, 1991]{beck91}
 Becker, \etal, 1991, {\it Astrophys. J. Suppl.} 75, 1.
\bibitem[Chiang \etal, 1995]{chia95}
 Chiang, J., C.E. Fichtel, C. von Montigny, P.L.Nolan and V. Petrosian, 
1995, \capj 452, 156.
\bibitem[Dermer and Gehrels, 1995]{derm95a}
 Dermer, C.D. and Gehrels, N., 1995, \capj 447, 103.
\bibitem[Dermer, 1995]{derm95b}
 Dermer, C.D., 1995, \capj 446, L63.
\bibitem[Dunlop and Peacock, 1990]{dunl90}
 Dunlop, J.S. and Peacock, J.A. 1990, \mn  247, 19.
\bibitem[Fichtel, 1995]{fich95}
 Fichtel, C.E., 1995, {\it Proc 3rd Compton Observatory Symposium}, {\it 
Astron Astrophys. Suppl.}, in press. 
\bibitem[Fichtel, Simpson, and Thompson, 1978]{fich78}
 Fichtel, C.E., Simpson, G.A. and Thompson, D.J. 1978, \capj, 222, 833.
\bibitem[Jang and Miller, 1995]{jang95}
 Jang, M. and H.R. Miller 1995, \capj 452, 582.
\bibitem[Kappadath \etal, 1995]{kapp95}
 Kappadath, \etal\ 1995, in {\it Proc. 24th Int. Cosmic Ray Conf., Rome} 
2, 230.
\bibitem[Kniffen \etal, 1993]{knif93}
 Kniffen, D.A., \etal, 1993, \capj 411, 133.
\bibitem[K\"{u}hr \etal, 1981]{kuhr81}
 K\"{u}hr, H., \etal, 1981 , {\it Astron. Astrophys. Suppl.}  45, 367.
\bibitem[Ledden and O'Dell, 1985]{ledd85}
 Ledden and O'Dell, 1985, {\it Astron. J.}\ 298, 630.
\bibitem[McNaron-Brown \etal, 1995]{mcna95}
 McNaron-Brown, K., \etal\ , 1995, \capj 451, 575.
\bibitem[Mukherjee \etal, 1995]{mukh94}
 Mukherjee, R., \etal, 1995 \capj 441, L61.
\bibitem[Mukherjee \etal, 1995]{mukh95}
 Mukherjee, R., \etal\ , 1995, submitted to \capj.
\bibitem[Padovani \etal, 1993]{pado93}
 Padovani, P. Ghisellini, G. Fabian, A.C. and Celotti, A. 1993, {\it 
Mon. Not. Royal Astron. Soc.}, 260, L21.
\bibitem[Reich, \etal, 1993]{reic93}
 Reich, W. \etal, 1993 {\it Astron. Ap.} 273, 65.
\bibitem[Salamon and Stecker, 1994]{sala94}
 Salamon, M.H. and F.W.Stecker, 1994, \capj 430, L21.
\bibitem[Sreekumar \etal, 1995]{sree95}
 Sreekumar, P., \etal\ 1995, in preparation.
\bibitem[Stecker, 1989]{stec89}
 Stecker, F.W. 1989, {\it Nucl. Phys. B} (proc. Suppl), 10B, 93.
\bibitem[Stecker, deJager, and Salamon, 1992]{stec92}
 Stecker, F.W. De Jager, O.C. and Salamon, M.H. 1992, \capj 390, L49.
\bibitem[Stecker, Salamon, and Malkan, 1993]{stec93}
 Stecker, F.W., Salamon, M.H. and Malkan, 1993, {\it Ap.J. Letters}, 
410, L71.
\bibitem[Thompson and Fichtel, 1982]{thom82}
 Thompson, D.J. and Fichtel, C.E., 1982, {\it Astron. Ap.} 109, 352.
\bibitem[Thompson \etal, 1995]{thom95}
 Thompson, D.J., \etal, 1995, \capj {\it Suppl}, in press.
\bibitem[Trombka \etal, 1977]{trom77}
 Trombka, J.I., \etal, 1977, \capj 212, 925.
\bibitem[Valtaoja and Ter\"{a}sranta, 1995]{valt95}
 Valtaoja, E. and Ter\"{a}sranta, H. 1995, {\it Astron. and Ap.} 297, L13.
\bibitem[von Montigny \etal, 1995]{mont95}
 von Montigny, \etal, 1995, \capj 440, 525.
\bibitem[Wagner and Witzel, 1995]{wagn95}
 Wagner, S.J. and A. Witzel 1995, {\it Ann. Rev. Astron. Ap.} 33, 163.
\bibitem[Wall and Peacock, 1985]{wall85}
 Wall, J.V. and Peacock, J.A. 1985, {\it MNRAS}, 216, 173.
\end{thebibliography}
\end{document}